\documentclass[lettersize,journal]{IEEEtran}
\usepackage{amsmath,graphicx,tabularx,makecell,multirow}
\usepackage{amsfonts}
\usepackage{enumerate}
\usepackage{bm}
\usepackage{algorithm} 
\usepackage{algorithmic} 
\usepackage{diagbox} 
\usepackage{mathrsfs}
\usepackage{stfloats}
\usepackage{enumerate}
\usepackage{epstopdf}
\usepackage{subfigure}
\usepackage{float}
\usepackage{amsbsy}
\usepackage{amsmath}
\usepackage{amssymb}
\usepackage{color}
\usepackage{ntheorem}
\usepackage{cite}

\theorembodyfont{\upshape}
\theoremheaderfont{\rmfamily\itshape}
\theoremseparator{:}
\theoremstyle{remark}

\begin{document}
\IEEEoverridecommandlockouts

\title{{Sparse MIMO for ISAC: New Opportunities and Challenges}\vspace{-1pt}}
\author{\IEEEauthorblockN{Xinrui~Li, Hongqi~Min, Yong Zeng, \emph{Senior Member, IEEE}, \\
Shi~Jin,  \emph{Fellow, IEEE}, Linglong~Dai, \emph{Fellow, IEEE}, Yifei~Yuan, \emph{Fellow, IEEE}, Rui~Zhang, \emph{Fellow, IEEE}\\
\thanks{X. Li, H. Min, Y. Zeng, and S. Jin are with the National Mobile Communications Research
Laboratory, Southeast University, Nanjing 210096, China. Y. Zeng is also
with the Purple Mountain Laboratories, Nanjing 211111, China (e-mail: \{xinrui\_li, minhq, yong\_zeng, jinshi\}@seu.edu.cn). (\emph{Corresponding author: Yong Zeng}.)}
\thanks{L. Dai is with Department of
Electronic Engineering, Tsinghua University, and also with
Beijing National Research Center for Information Science and
Technology (BNRist), Beijing 100084, China (e-mail: daill@tsinghua.edu.cn).}
\thanks{Y. Yuan is with Future Mobile Technology Lab, China Mobile Research
Institute, Beijing 100053, China (e-mail: yuanyifei@chinamobile.com).}
\thanks{R. Zhang is with School of Science and Engineering, Shenzhen Research Institute of Big Data, The Chinese University of Hong Kong, Shenzhen, Guangdong 518172, China (e-mail: rzhang@cuhk.edu.cn). He is also with the Department of Electrical and Computer Engineering, National University of Singapore, Singapore 117583 (e-mail: elezhang@nus.edu.sg).}
}}
\maketitle
\vspace{-1cm}
\begin{abstract}
Multiple-input multiple-output (MIMO) has been a key technology of wireless communications for decades. A typical MIMO system employs antenna arrays with the inter-antenna spacing being half of the signal wavelength, which we term as \textit{compact MIMO}.
Looking forward towards the future sixth-generation (6G) mobile communication networks, MIMO system will achieve even finer spatial resolution to not only enhance the spectral efficiency of wireless communications, but also enable more accurate wireless sensing. To this end, by removing the restriction of half-wavelength antenna spacing, \emph{sparse MIMO} has been proposed as a new architecture that is able to significantly enlarge the array aperture as compared to conventional compact MIMO with the same number of array elements. In addition, sparse MIMO leads to a new form of virtual MIMO systems for sensing with their virtual apertures considerably larger than physical apertures. As sparse MIMO is expected to be a viable technology for 6G, we provide in this article a comprehensive overview of it, especially focusing on its appealing advantages for integrated sensing and communication (ISAC) towards 6G. Specifically, assorted sparse MIMO architectures are first introduced, followed by their new benefits as well as challenges. We then discuss the main design issues
of sparse MIMO, including beam pattern synthesis, signal processing, grating lobe suppression, beam codebook design, and array geometry optimization. Last, we provide numerical results to evaluate the performance of sparse MIMO for ISAC and point out promising directions for future research.
\end{abstract}

\section{Introduction}
Since its theoretical breakthrough in 1990s, multiple-input multiple-output (MIMO) has become a critical technology for contemporary wireless systems and now evolved to the fifth-generation (5G) massive MIMO \cite{bjornson2019massive}. The standard MIMO system typically employs antenna arrays for which the  inter-antenna spacing is set to half of the signal wavelength, which we term as \textit{compact MIMO}. The main reasons for choosing half-wavelength as the inter-antenna spacing are three folds. Firstly, it is a separation without causing severe mutual coupling between adjacent array elements. Secondly, half-wavelength is roughly the channel coherence distance in rich scattering environment, so that setting antenna spacing to such a value can reap the full spatial diversity gain. Thirdly, half-wavelength is the maximum separation without causing grating lobes in the array's angular response.
Looking forward towards 2030s, the sixth-generation (6G) mobile communication networks need to provide super-dense connection with up to $10^8$ devices per $\rm km^2$, ultra-high peak data rate reaching to $1$ $\rm Tbps$, hyper-accurate positioning at the centimeter ($\rm cm$) level, and hyper-reliable service
of $99.99999\%$ reliability, by exploiting advanced channel coding, modulation, and transmission technologies \cite{1212,lee2023towards}. In addition, beyond traditional communication services, integrated sensing and communication (ISAC) is a new technology that jointly realizes wireless communication and
radar sensing functions in a unified system. In particular, with the evolution of millimeter-wave (mmWave) and terahertz (THz) communications, the frequency bands of
 communications gradually chime with those of radar sensing, which
 facilitates the effectuation of ISAC. Recently, ITU-R has identified ISAC as one of the six major usage scenarios for 6G \cite{1212, 10012421}. To achieve ambitious goals of 6G ISAC, future MIMO systems should be further advanced to achieve even finer spatial resolution to not only significantly enhance the spectral efficiency of wireless communications, but also enable more accurate wireless sensing for densely located targets. To this end, one straightforward approach is to significantly scale up the number of array elements, shifting from the existing 5G massive MIMO to future extremely large-scale MIMO (XL-MIMO) regime \cite{lu2023tutorial}. However, this will inevitably increase the system hardware, energy consumption, and signal processing costs. \par

A promising approach to address the above issue is {\it sparse MIMO}, which is an alternative MIMO architecture to achieve larger aperture than the conventional compact MIMO, without having
to increase the number of array elements. The key idea of sparse MIMO is to remove the restriction of half-wavelength antenna spacing, so that array elements can be more widely separated. In fact, the performance gain of sparse MIMO in aspects of wireless positioning and radar sensing has been extensively studied. For example, by applying (conjugate) correlation for signals of different sparse array (SA) elements, difference or sum co-arrays can be constructed, which can achieve virtual MIMO for sensing with considerably larger virtual apertures than physical apertures \cite{5456168, 5609222}. This thus significantly increases the sensing spatial resolution and degree of freedom (DoF). Note that in the context of localization and sensing, the aforementioned DoF represents the ability to sense the number of targets, which differs from the spatial multiplexing gain of MIMO communications. However, the method of signal (conjugate) correlation for virtual MIMO construction is difficult to be applied to communication systems. This is because different from sensing systems of which the main objective is to estimate the channel parameters like the path angles, communication systems need to decode information-bearing symbols, while the direct signal (conjugate) correlation processing leads to multiplication of communication symbols, which makes the symbol detection a more difficult task. It is worth mentioning that even without forming virtual arrays, it has been shown that sparse MIMO is still able to achieve significant performance gains over compact MIMO with the same number of array elements for both far-field and near-field communications \cite{wang2023can, li2023multi}, especially for densely located users. Specifically, as revealed in \cite{wang2023can}, while sparse MIMO generates undesired grating lobes, it achieves a narrower main lobe. The fact that the users' spatial frequency differences are typically non-uniformly distributed provides a natural filtering for the undesired grating lobes. This thus makes sparse MIMO outperform the conventional compact MIMO for communications \cite{wang2023can}. Despite of the promising benefits of sparse MIMO for both sensing and communications, to our best knowledge, a unified study on sparse MIMO for ISAC is still missing. This thus motivates this article to provide an overview of the new opportunities and challenges for 6G ISAC by shifting the MIMO design from conventional compact MIMO to sparse MIMO.\par
As an illustration, a basic sparse MIMO ISAC system with uniform sparse arrays (USAs) is shown in Fig. 1, where the base station (BS) communicates with a user equipment (UE) and simultaneously senses the location of a target. Both the BS and UE are equipped with USAs, and their inter-antenna spacings are respectively denoted by $\eta_{\rm BS}d_0$ and $\eta_{\rm UE}d_0$, where $\eta_{\rm BS}\geq 1$ and $\eta_{\rm UE}\geq 1$ are defined as the sparsity parameters, and $d_0=\frac{\lambda}{2}$ is the antenna
separation for conventional compact MIMO, with $\lambda$ being the signal wavelength. Note that for general sparse MIMO ISAC systems, the BS and UE can adopt different SA architectures as will be discussed in Section II.\par
The rest of this article is organized as follows. Assorted sparse MMO architectures
are introduced in Section II, followed by their new benefits and challenges for ISAC in Section III. Then, in Section IV, we discuss the main design issues of sparse MIMO, including its
beam pattern synthesis, signal processing, grating lobe suppression, beam codebook design, and array geometry optimization. In Section V, we numerically
demonstrate the superiority of sparse MIMO over compact
MIMO in terms of spatial resolution, sensing accuracy, and
spectral efficiency. Last, we conclude our article and point
out several promising directions for future research in Section VI.
\begin{figure}[t]
  \centering
    \includegraphics[scale=0.9]{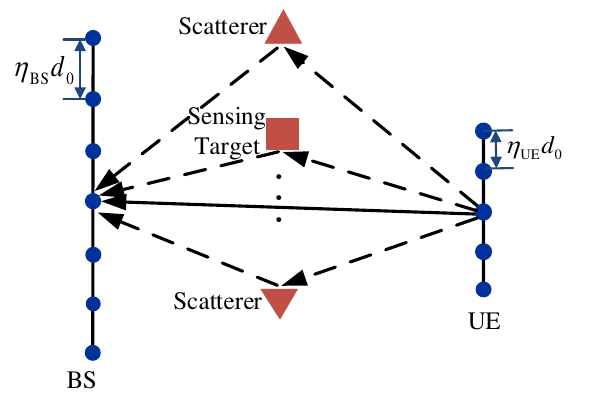}
  \caption{An illustration of sparse MIMO ISAC systems, where the BS and UE are equipped with USAs, with their inter-antenna spacings being $\eta_{\rm BS}d_0$ and $\eta_{\rm UE}d_0$, respectively. Note that for general sparse MIMO ISAC systems, the BS and UE can adopt different SA architectures as will be discussed in Section II.}\label{pic1}
  \label{12}
  \vspace{-0.3cm}
\end{figure}\par

\begin{figure*}[htp]
  \centering
    \includegraphics[scale=0.63]{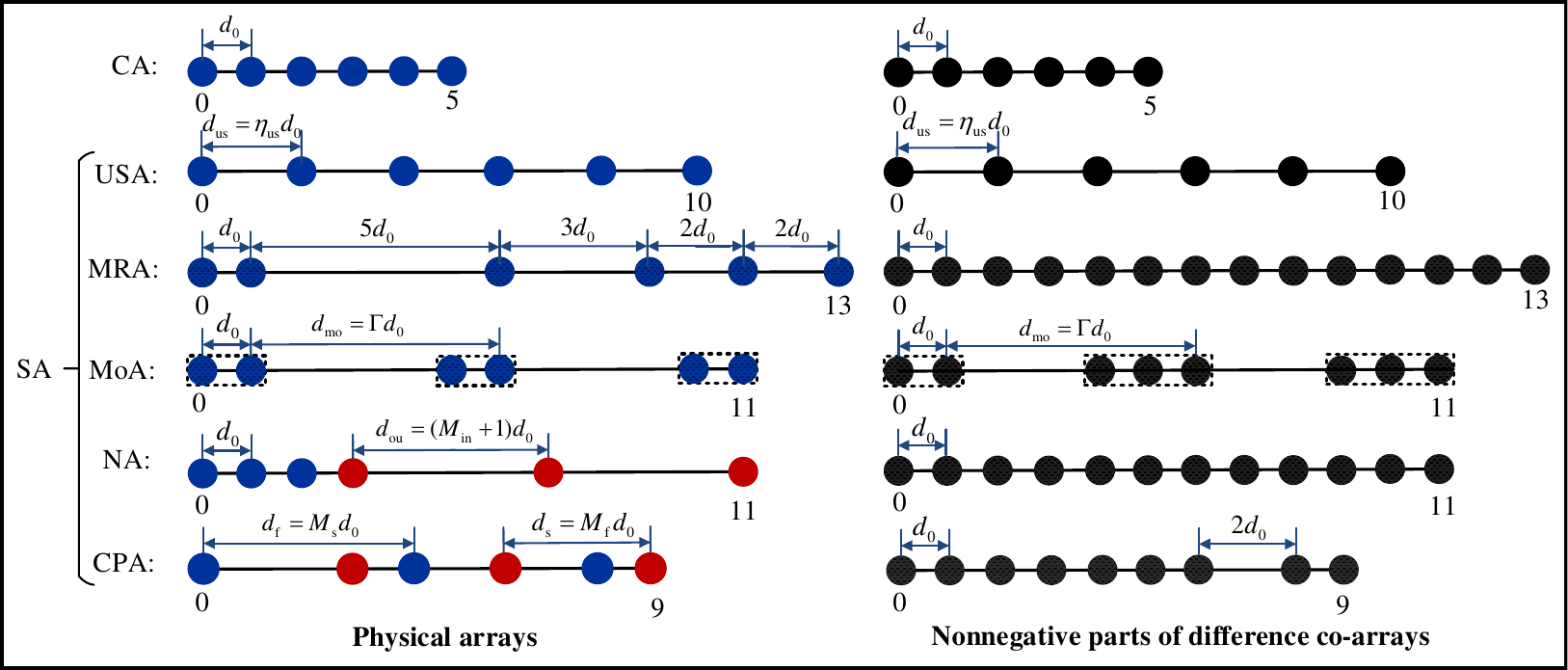}
  \caption{Illustration of different array architectures, all with $6$ elements, and their virtual arrays formed by the nonnegative parts of difference co-arrays. }\label{pic1}
  \label{12}
  \vspace{-0.3cm}
\end{figure*}\par
\section{Sparse MIMO Architectures}
 In general, depending on whether all adjacent elements are separated with an equal distance or not, the array architectures of sparse MIMO include USA \cite{wang2023can} and non-uniform sparse array (NUSA), elaborated as follows. \par

\textit{1) USA}: As shown in Fig. 2, for USA, all adjacent elements are separated with an equal distance, which is denoted as $d_{\rm us}=\eta_{\rm us}d_0$, where $\eta_{\rm us}\geq 1$ is the sparsity parameter of USA \cite{wang2023can}. Note that USA includes the conventional CA as a special case, by letting $\eta_{\rm us}=1$.\par
\textit{2) NUSA}: For NUSA, neighbouring array elements may have different separations. As seen from Fig. 2, typical examples of NUSA include minimum redundant array (MRA) \cite{4541350}, modular array (MoA) \cite{li2023multi}, nested array (NA) \cite{5456168}, and co-prime array (CPA) \cite{5609222}. It is also seen that in contrast to CA and USA, NUSAs can form larger continuous virtual apertures by using difference co-arrays. Due to the symmetric properties of difference co-arrays with respect to the zero point, we only show the virtual arrays formed by the nonnegative parts of difference co-arrays in Fig. 2 for brevity. More specifically, the geometries of various NUSAs are discussed below.\par

\textbf{MRA}: MRA uses the minimum
number of array elements to generate a maximum continuous virtual aperture, by minimizing duplicate pairs \cite{4541350}. For example, when the
number of  array elements is $6$, the set of inter-antenna spacings is $\{d_0,5d_0,3d_0,2d_0,2d_0\}$. Given a fixed number of array elements, MRA has the largest hole-free difference co-array among all the arrays shown in Fig. 2. However, there is no closed-form expression for such an array geometry, and finding the optimal element locations for MRA usually requires numerical search.\par

\textbf{MoA}: MoA is an architecture where the array elements are regularly mounted in a modular manner. Each of modules is an uniform array consisting of $M_{\rm mo}$ elements separated by half-wavelength, and there are $N_{\rm mo}$ modules that are separated with relatively large spacing, say $d_{\rm mo}=\Gamma d_0$, where $\Gamma\geq M_{\rm mo}$ is the inter-module spacing parameter \cite{li2023multi}.\par

\textbf{NA}: NA is obtained by concatenation of inner and outer arrays, where the inner array has $M_{\rm in}$ antenna elements with spacing equal to half-wavelength, and the outer array has $M_{\rm ou}$ antenna elements with spacing set to $d_{\rm ou}=(M_{\rm in}+1)d_0$ \cite{5456168}.
Different from MRA, NA has the closed-form expression for array element positions. \par

\textbf{CPA}: CPA is constructed by two USAs placed collinearly, where the first USA has $M_{\rm f}$ elements with the inter-antenna spacing $d_{\rm f}=M_{\rm s}d_0$, and the second USA has $M_{\rm s}$ elements with the inter-antenna spacing $d_{\rm s}=M_{\rm f}d_0$ \cite{5609222}. Note that $M_{\rm f}$ and $M_{\rm s}$ are co-prime in order to achieve high sensing DoF. \par

Besides the aforementioned sparse MIMO architectures, adaptive array architecture can be designed based on beamforming optimization in terms of signal-to-noise ratio (SNR) maximization, especially when distributions of sources are a priori known \cite{9216481}.\par

\section{New Opportunities and Challenges of Sparse MIMO for ISAC}
Without increasing the number of array elements, the physical aperture of compact MIMO is fundamentally  restricted by half-wavelength antenna spacing. In ISAC systems, this makes it difficult to further enhance spatial resolution, sensing DoF, as well as spatial multiplexing gain. As a comparison, thanks to the flexible inter-antenna spacings of various SAs as presented above, sparse MIMO for ISAC has several appealing new advantages:\par
\textit{1) Finer Spatial Resolution}: For far-field MIMO systems, spatial resolution is determined by the main lobe beamwidth in the angular domain, which is inversely proportional to the total array aperture \cite{li2023multi}. Thanks to its larger total array aperture, the angular resolution of sparse MIMO is finer than its compact MIMO counterpart with the same number of array elements. The improvement of spatial resolution is highly desirable for ISAC. On one hand, it reduces the inter-user interference (IUI), especially in hot spot areas where massive crowded users need to communicate simultaneously. On the other hand, the improvement of spatial resolution is conducive to distinguishing densely located sensing targets, and their estimated location information can be further used to assist in communication design.\par
\textit{2) Larger Sensing DoF}:
In radar systems, the sensing DoF is defined as the maximum number of targets that can be distinguished  simultaneously, which is dependent on the number of elements within the largest consecutive virtual array. By using $M$ physical array elements, sparse MIMO with NUSA architectures like MRA, NA or CPA, can achieve difference or sum co-arrays that have $O(M^2)$ virtual elements \cite{4541350, 5456168, 5609222}. Note that such difference or sum co-arrays can be implemented by signal (conjugate) correlation. This implies that a sensing DoF in the order of $O(M^2)$ can be achieved by sparse MIMO, as opposed to $O(M)$ for the conventional compact MIMO.

\textit{3) Enlarged Near-Field Region}: Near-field region corresponds to the scenario when the array aperture is no longer negligible as compared to the link distance, so that the conventional uniform plane wave (UPW) model becomes invalid. Instead, the spherical wavefront across array elements needs to be considered \cite{lu2023tutorial}. For the same number of antenna elements, sparse MIMO
will have larger near-field region than its compact counterpart, due to its larger total array apertures. This provides new opportunities for ISAC. For wireless communications, near-field MIMO not only leads to enhanced spatial multiplexing gain than its far-field counterpart, but also offers the additional distance dimension to suppress the IUI \cite{lu2023tutorial}. For radar sensing, the enlarged near-field region renders it possible to sense not only the target direction, but also its distance by using one single BS \cite{10388218}.\par

\textit{4) Reduced Mutual Coupling}: With the inter-antenna spacing larger than half-wavelength, sparse MIMO can reduce the electromagnetic coupling between antennas, which mitigates the effect of mutual coupling. On one hand, the low coupling effect may weaken beam direction offset, and thus achieves more accurate channel estimation for communications. On the other hand, it may stabilize the polarization state of electromagnetic wave, which improves the radar's ability to identify and classify various targets.\par

\textit{5) More Flexible Deployment}: With the restriction of half-wavelength antenna spacing, the conventional compact MIMO typically requires  a contiguous deployment platform, which is difficult to achieve in many scenarios. By contrast, sparse MIMO can be designed for conformal and flexible deployment in practice. For instance, MoA with relatively large inter-module spacing can be mounted on the building facades with adjacent modules separated by windows \cite{li2023multi}.\par

\textit{6) Saving of Hardware, Energy, and Signal Processing Cost}: To achieve the same total array aperture as its compact counterpart, sparse MIMO requires less activated antennas. This implies that for antenna deployment at the BS, sparse MIMO can not only achieve lightweight hardware, but also save the energy and signal processing costs \cite{9331675}.\par
However, compared to conventional compact MIMO, sparse MIMO for ISAC also faces new challenges:\par
\textit{1) Grating Lobes}: For sparse MIMO, undesired grating lobes are generated since its inter-antenna spacing is larger than half-wavelength. In particular, for USA, the amplitude and beamwidth of grating lobes are equal to that of main lobe. When users or targets are located in grating lobes, it induces severe IUI for communications and angular ambiguity for radar sensing.

Fortunately, in radar sensing systems, sparse MIMO with NUSA architectures like MRA, NA or CPA, has been designed to effectively circumvent undesired grating lobes by forming virtual arrays with more consecutive elements, which are based on difference or sum co-arrays implemented by signal (conjugate) correlation.
However, how to apply such techniques to communication or ISAC systems is still unclear.\par
\textit{2) Beam Split}: For far-field wideband ISAC, the beams over different frequencies may be split
into distinct directions. Compared to compact MIMO with the same number of array elements, sparse MIMO with a larger array aperture may exhibit a more significant beam split effect. When it comes to sparse XL-MIMO, the near-field beams over different frequencies are split into various physical locations, resulting in irregular split patterns induced by grating lobes. This leads to more difficult signal processing. \par

\section{Main Design Issues for Sparse MIMO ISAC}
In this section, we present the main design issues for sparse MIMO ISAC, including its beam pattern synthesis, signal processing, grating lobe suppression, beam codebook design, and array geometry optimization.\par
\subsection{Beam Pattern Synthesis}
The far-field beam pattern describes the array gain variation of a designed beam as a function of the spatial angle difference $\Delta_{\theta}\triangleq\sin \theta-\sin \theta_0$, where $\theta$ and $\theta_0$ are the observation direction and desired beamforming direction, respectively. To be specific, the far-field beam pattern of an $M$-element compact uniform linear array (ULA) is  $G_\mathrm{FF}^\mathrm{CA}(\Delta_\theta)=\frac{1}{M}\left|\frac{\sin\left(\frac{\pi}{2}M\Delta_{\theta}\right)}{\sin\left(\frac{\pi}{2}\Delta_{\theta}\right)}\right|$, with its angular resolution being $\frac{2}{M}$ \cite{li2023multi}. As a comparison, under the same number of array elements $M$, the far-field beam pattern of sparse ULA is $G_\mathrm{FF}^\mathrm{SA}(\Delta_\theta)=\frac{1}{M}\left| \sum_{m=0}^{M-1} e^{j\frac{2\pi}{\lambda} d_m\Delta_{\theta}} \right|$, where $d_m$ is the distance between the $m$-th antenna and the reference antenna, which depends on different sparse MIMO architectures discussed in Section II. In particular, \cite{wang2023can} and \cite{li2023multi} have derived the closed-form expressions for far-field beam patterns of sparse and modular ULAs, and their respective angular resolutions are $\frac{2}{\eta_{\rm us} M}$ and $\frac{2}{\Gamma N_{\rm mo}}$, which are finer than that of compact ULA since $\eta_{\rm us} \geq 1$ and $\Gamma\geq M_{\rm mo}$. However, in contrast to its compact counterpart, sparse MIMO may introduce undesired grating lobes due to the sparse sampling in the spatial domain.

Different from the far-field beam that aligns to a certain direction only, the near-field beam focusing pattern describes the array gain variation of a designed beam focused on the desired position as a function of the observation location. In \cite{lu2023tutorial} and \cite{li2023multi}, the closed-form expressions for near-field beam focusing patterns of compact, sparse and modular ULAs have been derived. It is revealed that in contrast to the compact counterpart with the same number of array elements, sparse and modular ULAs can provide finer spatial resolutions from both angular and distance dimensions \cite{lu2023tutorial,li2023multi}. Besides, compared to the far-field UPW model, the near-field spherical wavefront model endows sparse and modular ULAs the ability to suppress grating lobes due to the non-linear phase variations across array elements. The beam focusing patterns of other sparse MIMO architectures like NA, CPA and MRA, are presented via simulation results in Section V.
\subsection{Signal Processing}
Generally speaking, signal processing can be classified into far-field and near-field issues. In the far-field region of sparse MIMO, different NUSAs like MoA, NA and CPA, have been widely studied in radar and localization to enhance sensing capability by forming difference or sum co-arrays. The classic subspace-based algorithms, such as MUSIC or ESPRIT, are able to effectively estimate directions of arrival (DOAs). However, due to the non-singularity of covariance matrices, the subspace-based algorithms are invalid for estimation of coherent signals. Typically, CA and USA can adopt the spatial smoothing MUSIC algorithm for decorrelation, which, however, is no longer applicable for NUSAs since their antenna spacings are unequal. To solve this issue, one possible method is to consider the maximum-likelihood (ML) estimation algorithm since it is applicable to coherent signals.
Another method is to take into account the angular-domain sparsity, so that the compressed sensing (CS) scheme can be leveraged to estimate coherent signals' DOAs. \par

For sparse MIMO that has a large near-field region, new signal processing problems need to be solved, since the angular and distance parameters are naturally coupled in the polar domain. On one hand, by utilizing the channel sparsity in the polar domain,
the polar-domain orthogonal matching
pursuit (OMP) algorithm can be applied to the efficient estimation of
near-field channel parameters. On the other hand, it is possible to adopt a two-level estimation method that first estimates the angular parameters by exploiting the cross-correlation for targets/users, and then estimates the distance by using the MUSIC algorithm.

\subsection{Grating Lobe Suppression}
For sparse MIMO sensing, difference or sum co-arrays have been proven to effectively eliminate grating lobes to avoid angular ambiguity. However, this method cannot be applied to wireless communications since the direct correlation between antenna elements leads to multiplication of communication symbols.\par
Under this circumstance, one possible approach to address the grating lobe issue is via optimizing the positions of
antenna elements or leveraging the technique of movable antennas \cite{10286328}. Nevertheless, the flexible adjustment
of array elements is practically difficult to implement, and the dynamic movement of movable antennas is quite challenging to design.
Another effective approach is based on user grouping, so that users located within the ranges of grating lobes are allocated to different
time-frequency resource blocks for communications \cite{li2023multi}. Hence, in multi-user scenarios, user grouping offers an effective means to mitigate severe IUI among users caused by grating lobes.\par

\begin{figure*}[htp]
  \centering
  \subfigure{
    \label{1} 
    \includegraphics[scale=0.438]{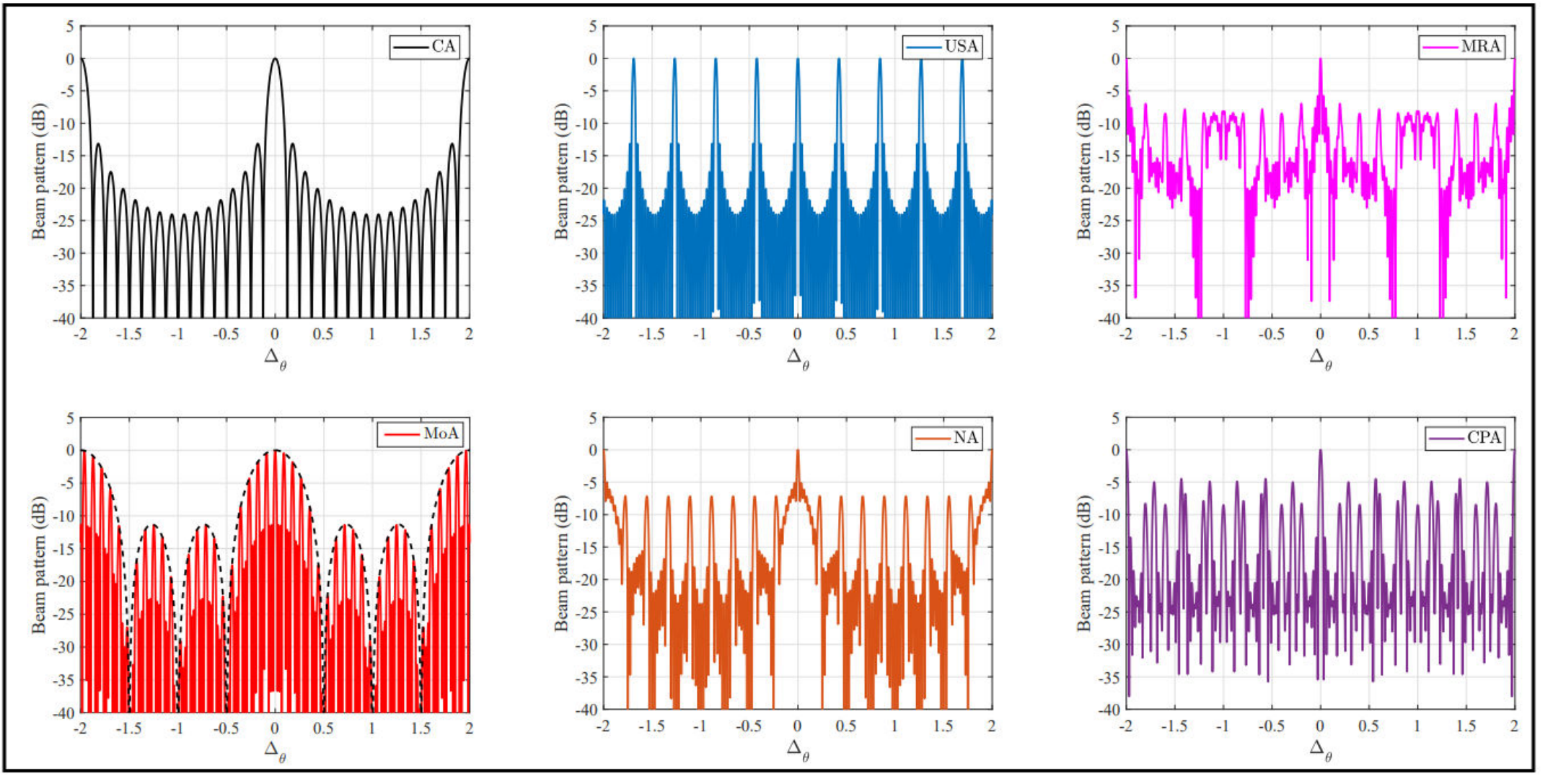}}
    \hspace{0in}
  \caption{The far-field beam patterns of different array architectures, all with $M=16$ antenna elements.}\label{pic1}
  \label{12}
  \vspace{-0.3cm}
\end{figure*}

\begin{figure*}[htp]
  \centering
  \subfigure{
    \label{1} 
    \includegraphics[scale=0.438]{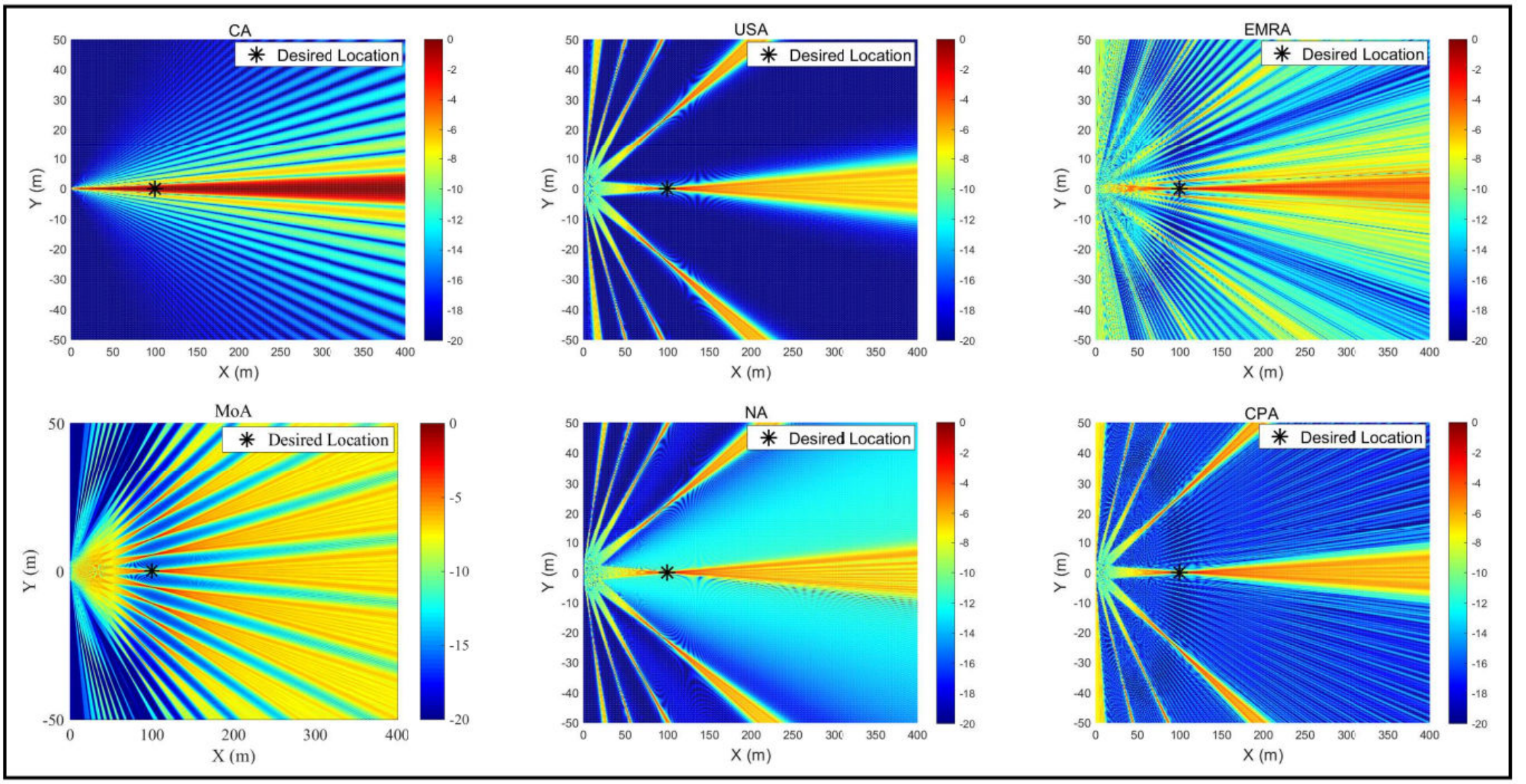}}
    \hspace{0in}
  \caption{The near-field beam focusing patterns of various array architectures, all with $M=128$ antenna elements. The desired beam focusing
location is $(r_0,\theta_0)=(200 \ \rm m,0^o)$, labelled as a black asterisk in the figure.}\label{pic1}
  \label{12}
  \vspace{-0.3cm}
\end{figure*}

\subsection{Beam Codebook Design}
Compact MIMO typically adopts the discrete fourier transform (DFT)-based beam codebook to form orthogonal beam codewords. However, sparse MIMO with NUSA architectures breaks the orthogonality of the conventional DFT-based beam codebook. In this case, the DFT-based beam codebook for sparse MIMO should be constructed by properly hollowing out partial elements of compact MIMO with the same array aperture.
Nevertheless, the DFT-based beam codebook for sparse MIMO may generate undesired grating lobes which cause severe IUI.\par
To alleviate the IUI, it is desirable to develop a more efficient codebook for sparse MIMO via optimization based on the minimization of side lobe level (SLL) or the maximization of peak-side lobe level ratio (PSLLR). Through the beamforming optimization, grating and side lobes with high levels can be more effectively suppressed in the spatial domain, but it may spread beam energy into unwanted directions or locations. If the user distribution is a priori known, the beam codebook can be designed by beamforming optimization. On one hand, the optimized beam codebook only needs to cover
the range of the given user distribution, which saves the number of codewords and thus accelerates the process of beam training. On the other hand, it is able to suppress the grating and side lobes for achieving a higher SINR.
To further accelerate the process of beam training, one possible approach is to take advantage of grating lobes to form multiple beams, so that scanning any section among adjacent lobes
can fully cover all possible directions or locations of users.\par

\subsection{Array Geometry Optimization}
Besides the typical sparse MIMO architectures mentioned above, the design of array geometry optimization according to practical applications is an important topic to explore as well. To achieve this goal, optimization techniques, such as convex optimization, particle swarm or machine learning, have been widely employed to optimize the positions of antenna elements, so as to reshape beam pattern, and correct beam direction.
These methods contribute to enhancing the capabilities of anti-interference and target sensing.
However, for sparse MIMO ISAC, it requires more advanced array geometry architectures to achieve a balance between communication and sensing performances. For instance, the authors in \cite{alidoustaghdam2022multibeam} developed a novel geometry architecture of sunflower array in ISAC systems, with its
phase centers determined by the maximum entropy principle, and sub-array elements selected by leveraging the geometry optimization.\par
\section{Simulation Results}

In this section, we numerically demonstrate the performance superiority of sparse
MIMO over compact MIMO in terms of spatial resolution, sensing accuracy, and
spectral efficiency.
Unless otherwise specified, the carrier frequency of mmWave is set to $f_0=28$ $\rm GHz$.
For fair comparisons, all array architectures are assumed to share the same number of array elements, i.e., $M=16$ in far-field and $M=128$ in near-field scenarios. Note that if the total aperture is fixed, sparse MIMO requires fewer antennas as compared to compact MIMO. For example, by fixing the array aperture as ${\rm A}_{\rm ref}=0.6858$ $\rm m$, CA requires $128$ antenna elements, while USA only needs $32$ antenna elements when its inter-antenna separation is given by $d_{\rm us}=4.1 d_0$.

Fig. 3 plots the far-field beam patterns of different array architectures. It is observed that compared to its compact counterpart with the same number of array elements, all array architectures of sparse MIMO provide finer angular resolution, which is expected since they have larger array apertures. Additionally, for the same array aperture, MoA generates more grating lobes than USA, since its inter-module spacing is larger than that of USA. It is also observed that grating lobes generated by USA have the same magnitude as the main lobe, which may cause the severe interference or angular ambiguity. By contrast, NUSAs, including MRA, MoA, NA, and CPA, can suppress the amplitude of undesired grating lobes to a certain extent, benefiting from their non-uniform array geometries.\par

Fig. 4 plots the near-field beam focusing patterns of different array architectures. Note that due to the prohibitive complexity of finding the optimal array element locations for MRA with $M=128$ elements, extended minimum redundancy array (EMRA) is considered as an alternative, which consists of $8$ sub-arrays, each with a $16$-element MRA \cite{4541350}.
It is observed that in contrast to its compact counterpart, all array architectures of sparse MIMO introduce grating lobes that may incur undesired ambiguous locations, while exhibiting more obvious near-field beam focusing effects on the desired locations for sensing/localization. It is also observed that the near-field beam focusing pattern can provide even finer spatial resolutions from both angular and distance dimensions, as compared to the far-field beam pattern only exhibiting angular resolution. These observations verify the
superiority of sparse MIMO over compact MIMO in improving the spatial resolution, especially in the near-field hot spot region.\par

Furthermore, we consider an uplink near-field sparse MIMO ISAC system, where $K=30$ users are uniformly scattered in a disk area, and the sensing target is located at $(r_t,\theta_t)=(200 \ \rm m, 70^o)$. The comparisons of near-field ISAC performances with different array architectures are presented in the following. Fig. 5 plots the normalized root mean squared error (NRMSE) of estimated target's DOAs versus the receive SNR, by using the zero forcing (ZF) based MUSIC algorithm.
It is observed from Fig. 5 that when the receive SNR is lower than $0$ $\rm dB$, NUSAs can sense more accurate DOA information than both USA and CA with the same number of array elements. This is expected since compared to the corresponding compact architecture, sparse MIMO with larger physical apertures is able to provide finer spatial resolution, whereas its induced grating lobe issues can be mitigated to some extent by near-field beamforming. Moreover, different from CA and USA, NUSAs can also form extra virtual elements with difference co-arrays, which contribute to greatly suppressing grating lobes. It is also observed that as the receive SNR increases beyond $0$ $\rm dB$, the sensing performances of all array architectures become quite good  by applying the MUSIC algorithm that is more suitable for high SNRs.

Fig. 6 plots the achievable sum rate versus the circular radius of user distribution, with far-field/near-field maximum ratio combining (MRC) beamforming, as well as the optimized user grouping \cite{li2023multi}. As can be seen from Fig. 6, for near-field MRC beamforming, when the circular radius of user
distribution is relatively small, the achievable sum rates of sparse MIMO are much larger than compact MIMO due to their finer spatial resolutions from both angular and distance dimensions. When the circular radius of user
distribution becomes large enough, the achievable sum rates of all array architectures tend to be similar. Furthermore, it is shown that the practical near-field beamforming achieves significantly higher spectral efficiency than the far-field beamforming.\par
\begin{figure}[t]
      \centering
    \label{1} 
    \includegraphics[scale=0.6]{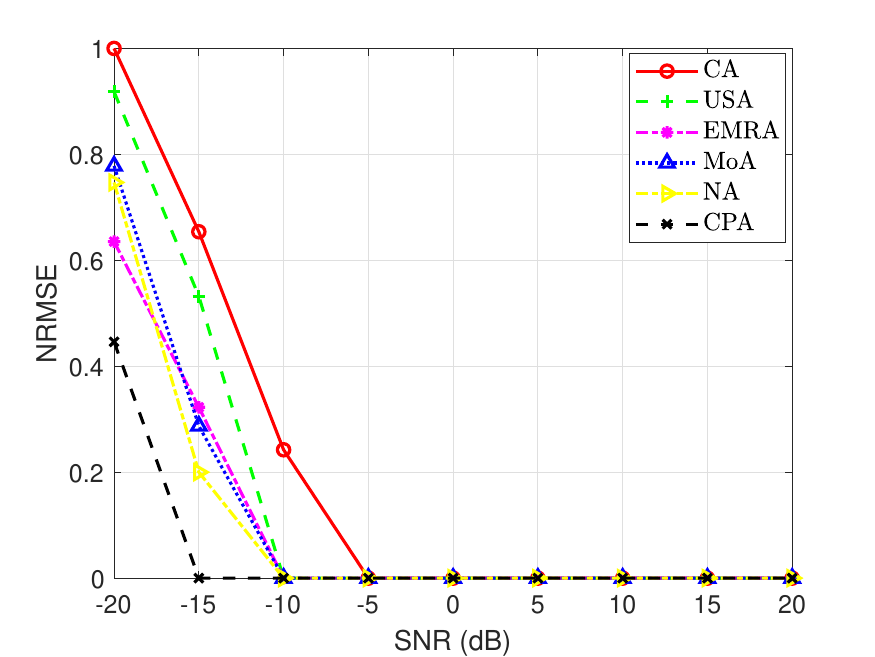}\vspace{-0.4cm}
    \hspace{0in}
  \caption{NRMSE of estimated target's DOAs versus the receive SNR, by using the ZF-MUSIC method.}\label{pic1}
  \label{12}
  \vspace{-0.4cm}
\end{figure}

\begin{figure}[t]
    \centering
    \label{1} 
    \includegraphics[scale=0.6]{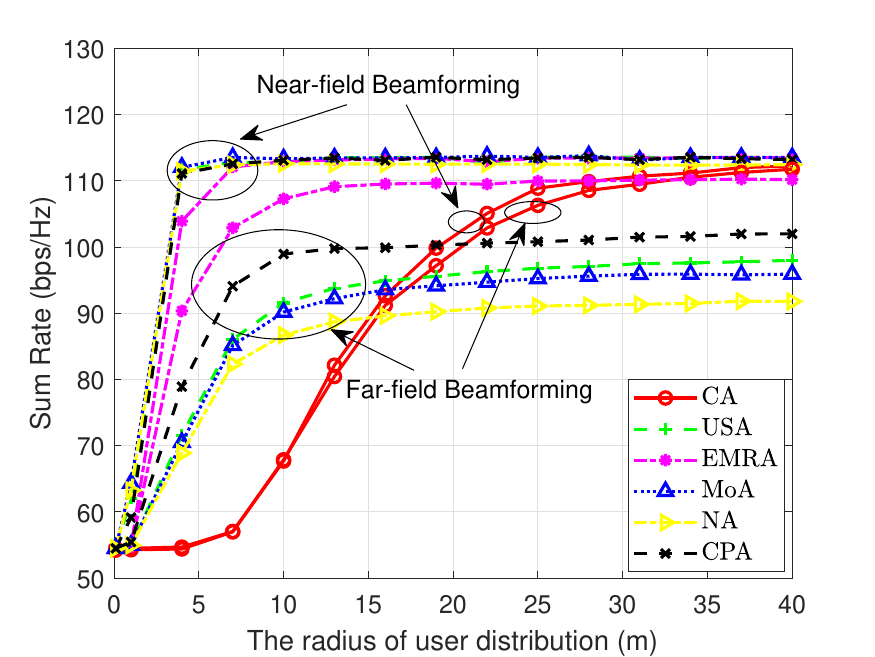}\vspace{-0.4cm}
    \hspace{0in}
  \caption{Achievable sum rate versus the circular radius of user distribution, with far-field/near-field MRC beamforming, and optimized user grouping.}\label{pic1}
  \label{12}
  \vspace{-0.5cm}
\end{figure}
\section{Conclusions and Future Directions}\label{g} \vspace{-1pt}
In future 6G ISAC systems, to achieve even finer spatial resolution, increasing the number of array elements with the conventional compact MIMO inevitably leads to further growing hardware,
energy, and signal processing costs. Instead, a more cost-effective approach is to use sparse MIMO, which removes the restriction
of half-wavelength antenna spacing to enlarge both the physical and virtual array apertures.
The practical architectures, new benefits and challenges, as well as main design issues of sparse MIMO have been discussed in this article, while more research efforts are needed for future investigation, particularly along the following directions.\par
 \textit{1) Sparse intelligent reflecting surface (IRS)/reconfigurable intelligent surface (RIS)}: In contrast to the active antenna array, the passive reflecting elements of IRS/RIS require a much larger array aperture to perform well in practice. In this case, sparse IRS/RIS is able to achieve a large array aperture cost-effectively, and makes it possible for conformal deployment on building surfaces. Moreover, to achieve even higher
spatial diversity gains with fewer elements, it is possible to consider movable sparse IRS/RIS, which enables the local movement of reflecting/reconfigurable elements.\par
  \textit{2) Beam control and tracking}: Sparse MIMO with large inter-antenna spacings has the potential to form multiple non-coherent beams pointing into different directions or locations simultaneously, which enables fast beam tracking in high-mobility scenarios. For example, NA can use the inner array to form wide beams to track fast-moving users/targets without frequent beam switching, while controlling the outer array to form sharp beams to point towards quasi-static users/targets for higher data rates or sensing accuracy.\par
  \textit{3) Physical layer security}: Benefiting from its larger array aperture as compared to compact MIMO with the same number of array elements, sparse MIMO is able to provide finer spatial resolution for sensing potential eavesdroppers more accurately based on their echo/reflected signals. With the  accurate locations of eavesdroppers obtained, the secrecy communication rate can be enhanced by jointly applying the  beamforming and artificial noise techniques.
\bibliographystyle{IEEEtran}
\bibliography{ref}
\end{document}